# Crossover between Anomalous Peak Effects Induced by Splayed and Tilted Columnar Defects in Ba$_{0.6}$K$_{0.4}$Fe$_2$As$_2$

Kanta Kato[1], Ryosuke Sakagami[1], Satoru Okayasu[2], Ataru Ichinose[3] and Tsuyoshi Tamegai[1]*

[1]Department of Applied Physics, The University of Tokyo, Hongo, Bunkyo-ku, Tokyo, 113-8656, Japan

[2] Advanced Science Research Center, Japan Atomic Energy Agency, Tokai, Ibaraki 319-1195, Japan

[3] Grid Innovation Research Laboratory, Central Research Institute of Electrical Power Industry, Yokosuka, Kanagawa 240-0196, Japan

We investigated the magnetic field dependence of the critical current density ($J_c$) in Ba$_{0.6}$K$_{0.4}$Fe$_2$As$_2$ with various configurations of columnar defects (CDs) introduced by 2.6 GeV U or 320 MeV Au irradiations. Splayed CDs are introduced by crossing CDs at a specific angle with respect to the $c$-axis, while tilted CDs are introduced by irradiating the sample from the direction tilted from the $c$-axis. We also prepared samples with asymmetric splayed CDs, which bridge the splayed CDs and the tilted CDs, by starting from the tilted CDs and adding CDs along the symmetric direction with respect to the $c$-axis. In all cases, non-monotonic magnetic field dependence of $J_c$, which we call anomalous peak effect, is observed at some fraction of the matching field when the magnetic field is applied along the specific direction depending on the configuration of CDs. We propose a model that explains the behavior of the anomalous peak effect in samples with different configuration of CDs.

1. **Introduction**

Motion of quantized vortices in superconductors under the magnetic field generates energy dissipation and limits the maximum current that can flow in superconductors. Defects in superconductors can suppress the motion of vortices effectively. Introduction of artificial defects by particle irradiations have been known to be a useful tool to enhance the critical current density ($J_c$) in superconductors. In particular, heavy-ion irradiation can create linear amorphous tracks in superconductors, which are called columnar defects (CDs) having similar geometry and dimensions to vortices and is considered to be the most effective way to enhance $J_c$. Significant enhancement of $J_c$ by the introduction of CDs has been demonstrated in cuprate superconductors,[1-6] iron-based superconductors (IBSs),[7-12] and conventional superconductors such as NbSe$_2$.[13-15]

It has been proposed that further enhancement of $J_c$ is achieved by dispersing the direction of the CDs.[16] A study on YBa$_2$Cu$_3$O$_{7-\delta}$ indirectly reported the effectiveness of splayed CDs by showing the correlation between the degree of natural splay, namely, misalignment of the damage tracks and $J_c$.[17] More direct evidence for the enhancement of $J_c$ by splayed CDs was presented in subsequent studies not only in YBa$_2$Cu$_3$O$_{7-\delta}$ but also in other cuprates superconductors[18-21] and IBSs.[22]

Enhancing $J_c$ stands as a critical objective in superconductor development for practical applications. Extensive efforts have focused on fabricating superconducting wires capable of sustaining large $J_c$ to generate extremely high magnetic fields without energy dissipation. A thorough understanding of vortex pinning mechanisms is essential for this endeavor. $J_c$ usually decreases monotonically as the applied magnetic field increases. However, an anomalous peak effect (APE), where $J_c$ shows a peak in a certain field range, is observed in superconductors with splayed CDs, which are formed by crossing CDs at a specific angle symmetrically with respect to the $c$-axis.[23] There are some features of this phenomenon. First, this anomalous non-monotonic behavior is generally observed in the magnetic field ($H$) region of about 1/3 of the matching field $B_\Phi$.[23] Second, APE occurs when the external field is applied along the average direction of the splayed CDs. When the applied field deviates from the average direction of the splayed CDs, APE is gradually suppressed.[23-25] Although a similar APE has also been observed in YBa$_2$Cu$_3$O$_{7-\delta}$ with tilted CDs at $H \sim B_\Phi/3$ when the field is applied parallel to CDs,[2] the presence of natural splay of CDs due to low energy Sn ions makes it possible to interpret the APE is caused by the splay of CDs. So, APE has been considered to be a unique phenomenon observed only in superconductors with splayed CDs.

However, a similar non-monotonic behavior of $J_c$ was reported in a recent study on 1.4 GeV Pb-irradiated NbSe$_2$ with CDs tilted from the $c$-axis by 30º without natural splay, when the magnetic field is applied along the CDs.[13] The authors attributed the broad peak structure observed in the magnetization curve to the self-field peak. The self-field peak is a peak effect that occurs near the self-field in superconductors with CDs parallel to the $c$-axis for $H//$CDs due to the bending of vortex lines at fields lower than the self-field, which makes the pinning by CDs weak near zero field.[9] However, the peak position reported in that paper is about 8 times larger than the self-field, which makes their interpretation questionable. On the other hand, the observed peak field is about $B_\Phi/5$, which is close to APE observed in systems with splayed CDs. If APE is observed in superconductors with tilted CDs, we may need to reconsider the mechanism of APE. This is because, although the detailed origin of APE is not fully understood, it was supposed that the entanglement of vortices trapped in splayed CDs was involved in the occurrence of APE.

In the present study, we aimed at clarifying not only whether APE appears in IBSs with tilted CDs but also the conditions and mechanism of APE. To this end, we irradiated Ba$_{0.6}$K$_{0.4}$Fe$_2$As$_2$ with 2.6 GeV U ions or 320 MeV Au ions to prepare samples with various configurations of CDs, starting from the tilted CDs and by adding CDs with various densities in the other symmetric direction to the tilted CDs. We studied the evolution of the magnetic field dependence of $J_c$ by changing the direction of the magnetic field. Based on the observed behavior of $J_c$ in samples with various configurations of CDs, we propose a model that explains the behavior of the peak field.

## 2. Experimental Methods

Ba$_{0.6}$K$_{0.4}$Fe$_2$As$_2$ single crystals ($T_c \sim 38$ K) were synthesized by FeAs self-flux method.[26,27] The crystals were shaped into rectangular parallelepipeds with thickness less than 10 μm, which is thinner than the projected range of 2.6 GeV U ions (62.6 μm) and 320 MeV Au ions (16.7 μm) in the crystal. CDs were installed in the crystal by 2.6 GeV U irradiation or 320 MeV Au irradiation. Irradiation of 2.6 GeV U was performed using the RIKEN Ring Cyclotron, and irradiation of 320 MeV Au was performed using tandem accelerator in JAEA. The incident direction of ions was changed by tilting the sample holder with the crystals about their *ab*-plane, and the angle of CDs ($\theta_{CD}$) is defined by the angle between their *c*-axis and the ion beam. The irradiation dose is evaluated by the dose-equivalent magnetic field, $B_\Phi$, called "matching field", at which all CDs are occupied by single vortices.

$$B_\Phi = n\Phi_0. \quad (1)$$

Here, $n$ is the areal density of CDs and $\Phi_0$ is the flux quantum.

Cross-sectional observations of the single crystals were performed with a STEM (JEM-2100F, JEOL). The spatial resolution of this microscope is 0.2 nm. However, we set it as 0.5 nm to add contrast in the STEM image. The specimen for STEM were prepared by digging and milling using a focused-ion beam (FIB), which is called the microsampling technique. The final milling using FIB was conducted at an acceleration voltage of 30 kV and with a sufficiently small ion current of 10 pA without tilting the specimen.

Magnetization of the crystal was measured by a superconducting quantum interference device (SQUID) magnetometer (MPMS-5XL, Quantum Design). Here, we define $\theta_H$ as the angle between the external magnetic field and the *c*-axis. To measure field-angle dependence of magnetization, we used an acrylic sample holder with a rotating sample stage as shown in Fig. S1 in supplemental materials.[28] Except for a sample with splayed CDs, all magnetization measurements were done at $T = 15$ K, where APE is the strongest. Average in-plane $J_c$ is calculated from the results of the magnetization measurements using the extended Bean model.[29,30] In the case of fields tilted from the *c*-axis ($\theta_H \neq 0°$), we corrected the magnetic field to $\widetilde{H}$ by using Blatter's scaling;[31]

$$\widetilde{H} = \varepsilon_\theta H \quad (2)$$

$$\varepsilon_\theta^2 = \varepsilon^2 \sin^2\theta_H + \cos^2\theta_H, \qquad (3)$$

where $\varepsilon$ ($< 1$) is the anisotropy parameter and we adopted $\varepsilon = 1/\gamma \sim 1/1.5$, the typical value for (Ba,K)Fe$_2$As$_2$ at 15 K.[32)] We also corrected $J_c$' calculated from the measured magnetization, which is the component parallel to the external field, to the real in-plane $J_c$;

$$J_c = J_c'/\cos\theta_H. \qquad (4)$$

This correction assumes that the supercurrent flows in the *ab*-plane and the direction of the magnetization is parallel to the *c*-axis in our thin samples for $\theta_H$ less than 45°. This assumption is confirmed by our previous study.[25)] We introduced various configurations of CDs with different defect densities on the two directions by irradiating the sample from two symmetric directions with respective to the *c*-axis ($\pm\theta_{CD}$). We first introduced tilted CDs with a dose of $B_\Phi^+ = 3$ T (U) or 4 T (Au) at an angle of $\theta_{CD} = 20°$ from the *c*-axis as shown in Fig. 1(a). Then, we added tilted CDs with different dose of $B_\Phi^-$ (0.1, 0.3, 0.5, 1, 2, and 3 T) in the direction at $\theta_{CD} = -20°$, which resulted in asymmetric splayed CDs as shown in Fig. 1(b). Finally, by making $B_\Phi^- = 3$ T (U) or 4 T (Au), we prepared a sample with (symmetric) splayed columnar defects as shown in Fig. 1(c).

## 3. Results and Discussion

STEM images of CDs created by 2.6 GeV U irradiation with $B_\Phi = 3$ T and 320 MeV Au irradiation with $B_\Phi = 4$ T + 1 T are shown in Fig. 2(a) and (b), respectively. CDs are nearly continuous in both cases, although the degree of continuity is better in the former case.

The magnetic field dependence of $J_c$ at several temperature in Ba$_{0.6}$K$_{0.4}$Fe$_2$As$_2$ with splayed CDs created by 2.6 GeV U and 320 MeV Au irradiation are shown in Fig. 3(a) and (b), respectively. In both cases, APE is observed between 5 K and 25 K. Peak fields are $H_p \sim 20$ kOe and $H_p \sim 25$ kOe for the 2.6 GeV U- and 320 MeV Au- irradiated samples, respectively. These values are about 1/3 of the total $B_\Phi$ in both cases.

The magnetic field dependence of $J_c$ at 15 K for different $\theta_H$ in Ba$_{0.6}$K$_{0.4}$Fe$_2$As$_2$ with splayed CDs created by 2.6 GeV U and 320 MeV Au irradiations are shown in Fig. 3(c) and (d), respectively. APEs at $H = B_\Phi/3$ are most pronounced when the applied field is aligned with the *c*-axis ($\theta_H = 0°$) and is suppressed as the applied field deviates from *c*-axis. When the applied field is parallel to one of the CDs, APE is suppressed completely.

The magnetic field dependence of $J_c$ at 15 K for different $\theta_H$ in Ba$_{0.6}$K$_{0.4}$Fe$_2$As$_2$ with tilted CDs created by 2.6 GeV U and 320 MeV Au irradiations are shown in Fig. 3(e) and (f), respectively. When the applied field is aligned with the CDs ($\theta_H = \theta_{CD} = 20°$), broad peaks show up at $H \sim B_\Phi/2$ in both cases. As the applied field deviates from the CDs, the peak position gradually shifts to lower fields. When the applied field is completely deviated from the CDs, such as at $\theta_H = -20°$, a small sharp peak remains at $H \sim 5$ kOe and $H \sim 2$ kOe for 2.6 GeV U- and 320 MeV Au-irradiated samples, respectively. These magnetic fields coincide with the calculated self-field of the U irradiated sample ($B_{sf} \sim J_c d \sim 4$ kG) and the Au irradiated sample ($B_{sf} \sim J_c d \sim 2.5$ kG), where $d$ is the thickness of the sample, indicating that it is a self-field peak. The peak fields at $\theta_H = 20°$ are $H_p \sim 15$ kOe and $H_p \sim 20$ kOe for U- and Au-irradiated sample, respectively. These peak fields are sufficiently large compared to their self-fields, indicating that they are due to APE.

Figure 4 (a) shows the magnetic field dependence of $J_c$ in Ba$_{0.6}$K$_{0.4}$Fe$_2$As$_2$ with asymmetric CDs introduced by 320 MeV Au with $B_\Phi^+ = 4$ T ($\theta_{CD} = 20°$) and $B_\Phi^- = 0.1$ T ($\theta_{CD} = -20°$). Hereafter, the sample with such irradiations is denoted by $B_\Phi = 4$ T + 0.1 T. By adding CDs as small as $B_\Phi^- = 0.1$ T, $J_c$ at the peak field is significantly increased from 4.5 MA/cm$^2$ ($B_\Phi = 4$ T) to 5.8 MA/cm$^2$ ($B_\Phi = 4 + 0.1$ T). This significant enhancement of peak $J_c$ suggests that the presence of intersections of CDs have a significant impact on APE. The peak becomes the strongest when the applied field is nearly parallel to the direction of CDs with higher density ($\theta_{CD} = 20°$). Figure 4 (b) shows the magnetic field dependence of $J_c$ in Ba$_{0.6}$K$_{0.4}$Fe$_2$As$_2$ with asymmetric CDs introduced by 320 MeV Au ($B_\Phi = 4$ T + 1 T, $\theta_{CD} = \pm 20°$). The peak is most pronounced at $\theta_H = 10$-$15°$, which is slightly shifted towards the $c$-axis compared with the direction of majority CDs at $\theta_{CD} = 20°$. In addition, the peak field is shifted to lower fields ($H_p \sim 17.5$ kOe) compared with the case of tilted CDs, despite the fact that the total $B_\Phi$ is increased. Magnetic field dependence of $J_c$ in other samples with different asymmetric splayed CDs are shown in Fig. S2 in supplemental materials.[33]

Figure 5(a) summarizes how the peak $\theta_H$ changes when the irradiation dose along $\theta_{CD} = -20°$ direction ($B_\Phi^-$) increases. The magnetic field angle where the APE shows up at the highest magnetic field is denoted as $\theta_H^{peak}$, and the average value is adopted when there are multiple maxima. $\theta_H^{peak}$ decreases monotonically as $B_\Phi^-$ increases. This behavior of $\theta_H^{peak}$ can be explained as follows. Figure 6(a) shows a schematic configuration of vortices and asymmetric splayed CDs. Note that Fig. 6(a) is a simplified schematic two-dimensional representation that ignores the thickness in the depth direction perpendicular to the splay plane. We assume that APE is the strongest when the internal magnetic induction along the two directions of CDs, $B^+$ and $B^-$ in Fig. 6(b), are proportional to the density of CDs for both directions, $B_\Phi^+$ and $B_\Phi^-$. In such a case, referring Fig. 6(b), the average direction vector $\boldsymbol{F}$ of vortices is given by the following equation;

$$\boldsymbol{F} = \begin{pmatrix} B_\Phi^+ \sin\theta_{CD} - B_\Phi^- \sin\theta_{CD} \\ B_\Phi^+ \cos\theta_{CD} + B_\Phi^- \cos\theta_{CD} \end{pmatrix}. \tag{5}$$

The angle that gives the strongest APE, $\theta_H^{peak}$ is expressed by the following formula;

$$\theta_H^{\text{peak}} = \frac{\pi}{2} - \arg(\boldsymbol{F}) = \arctan\left(\tan\theta_{\text{CD}} \cdot \frac{B_\Phi^+ - B_\Phi^-}{B_\Phi^+ + B_\Phi^-}\right). \tag{6}$$

A comparison of the actual data and the Eq. (6), shown as a dashed line in Fig. 5(a), demonstrates that this model accounts for the variation of $\theta_H^{\text{peak}}$ very well.

As we have described above, APE occurs at magnetic fields close to some fraction of $B_\Phi$. Figure 5(b) summarized how the peak field changes with $\theta_H$ in Ba$_{0.6}$K$_{0.4}$Fe$_2$As$_2$ with different configurations of CDs. The peak field in this figure is normalized by the total $B_\Phi$ ($= B_\Phi^+ + B_\Phi^-$) for each sample. $H_\text{p}/B_\Phi$ as a function of $\theta_H$ shows a sharp peak in the sample with tilted CDs, while the peak becomes broader in the sample with splayed CDs. The maximum value of $H_\text{p}/B_\Phi$ is ~1/2 in the sample with tilted CDs, while it approaches to ~1/3 in the sample with splayed CDs. In the system with asymmetric splayed CDs, $H_\text{p}/B_\Phi$ changes continuously from 1/2 to 1/3 as $B_\Phi^-$ along the symmetric direction opposite to the tilted CDs increases. As we observed, $\theta_H^{\text{peak}}$ and the peak magnetic field, which are important features of APE, show continuous changes as the CDs in the system changed from tilted CDs to splayed CDs. These facts may suggest that the origins of APEs in samples with splayed and tilted CDs are the same. However, it is reasonable to infer that APEs observed in samples with splayed CDs and tilted CDs originate from distinct mechanism by the reason described below.

To understand the origin of APE, it is important to note that APE is not observed in a sample with parallel CDs.[23] When APE is observed in a sample with splayed CDs, we suggested the importance of the crossing point of CDs as shown in Fig. 7(a).[24] Imagine a situation where two vortices are individually trapped on neighboring CDs pointing in different directions at a magnetic field close to but lower than $B_\Phi$. In order to move one vortex leaving another behind, the two vortices are forced to entangle. Large entanglement energy barrier can suppress vortex motion and can cause APE. However, in the case of tilted CDs, there are no such crossing points of CDs. Thus, suppression of vortex motion leading to APE due to a similar mechanism cannot work. Another important feature that we need to consider in the case of tilted CDs is the fact that vortices have to intersect the surface of the sample perpendicularly, since no superconducting current can flow outside of the sample. It immediately means that vortices near the surface cannot be pinned by tilted CDs, thus forming kinks near the surface as shown in Fig. 7(b). Unlike the entanglement of vortices in a system with splayed CDs, such vortex kinks near the surface are mobile and can trigger the vortex motion as a whole. However, when the density of vortex kinks increases at a magnetic field close to but lower than $B_\Phi$, kink-kink interactions may hinder the motion of kinks, generating stronger pinning of vortices as a whole, leading to the anomalous enhancement of $J_\text{c}$. Obviously these two scenarios for APE in systems with splayed CDs and tilted CDs are completely different.

## 4. Summary

We have studied the dependence of $J_c$ on applied magnetic field, in $Ba_{0.6}K_{0.4}Fe_2As_2$ single crystals with various configurations of CDs introduced by 2.6 GeV U and 320 MeV Au irradiations. In the system with symmetrically splayed CDs, APE is observed at around $H \sim B_\Phi/3$ when the magnetic field is applied parallel to the average direction of the splayed CDs ($c$-axis). On the other hand, in the system with purely tilted CDs, APE is observed at around $H \sim B_\Phi/2$ when the applied magnetic field is aligned to the direction of the CDs. Furthermore, in the system with asymmetric splayed CDs, peak $\theta_H$ and peak magnetic field changed continuously to take intermediate values between the values in the systems with symmetrically splayed CDs and tilted CDs, respectively. The variation of peak $\theta_H$ could be reproduced by a simple model assuming that APE is enhanced when vortices occupy each family of CDs in proportion to their densities. Despite the fact that some of the characteristics of APE in systems with splayed CDs and tilted CDs change continuously, detailed consideration on the elementary process of vortex dynamics suggests that they may originate from different mechanisms.

**Acknowledgment**

We thank A. Yoshida and T. Kambara for their help in the irradiation experiments in RIKEN.

*E-mail: tamegai@ap.t.u-tokyo.ac.jp

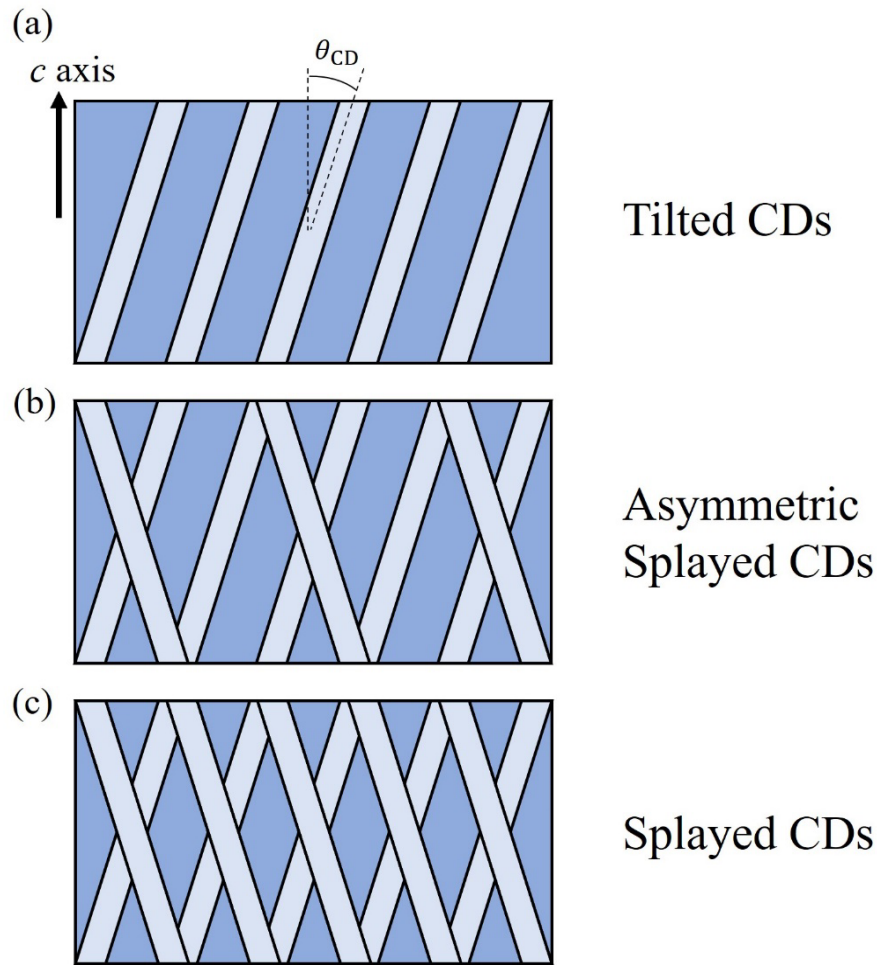

Fig. 1. (Color online) Schematic images of (a) tilted CDs, (b) asymmetric splayed CDs, and (c) splayed CDs.

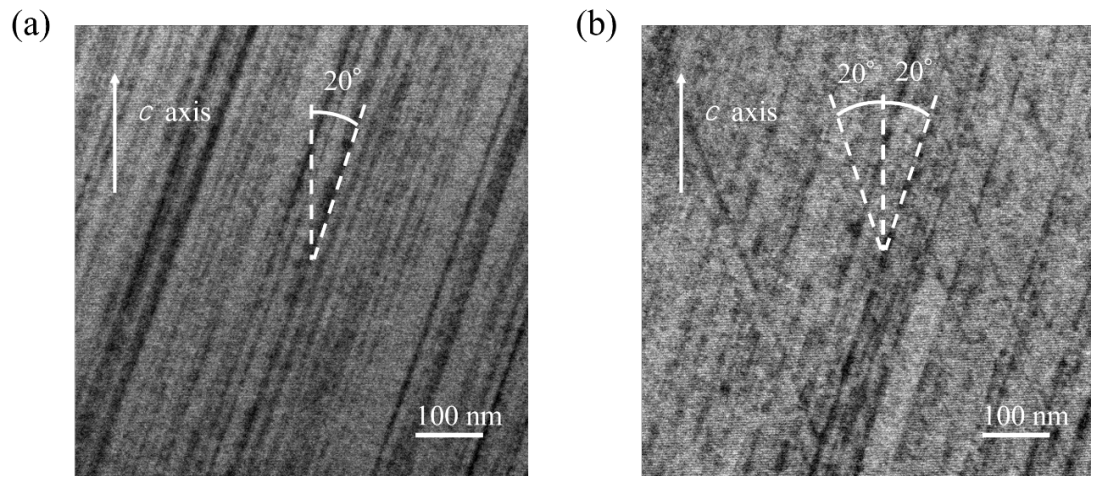

Fig. 2. (a) STEM image of cross-sections of 2.6 GeV U irradiated Ba$_{0.6}$K$_{0.4}$Fe$_2$As$_2$ ($B_\Phi$ = 3 T, $\theta_{CD}$ = 20°). (b) STEM image of cross-sections of 320 MeV Au irradiated Ba$_{0.6}$K$_{0.4}$Fe$_2$As$_2$ ($B_\Phi$ = 4 T + 1 T, $\theta_{CD}$ = ±20°).

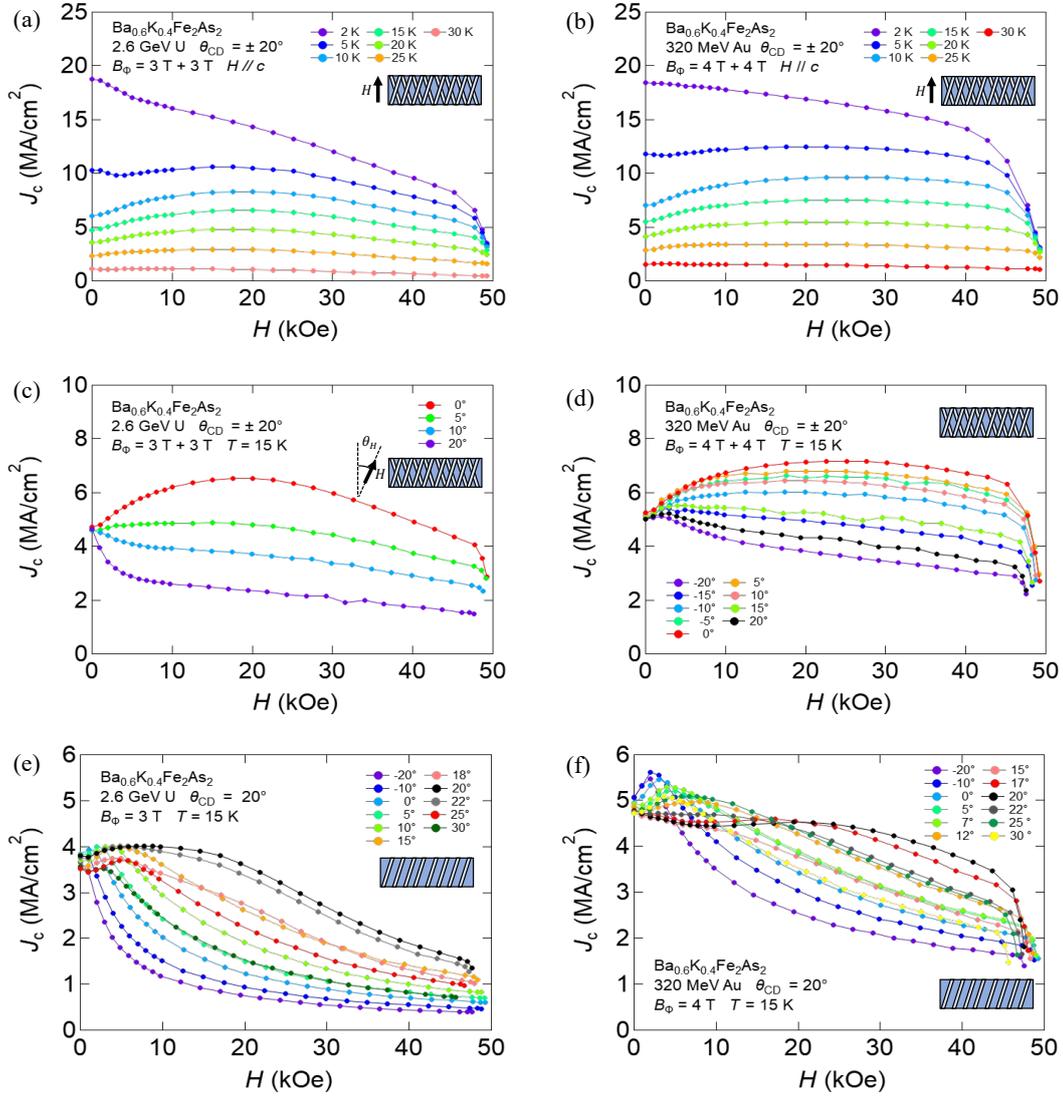

Fig. 3. (Color online) Magnetic field dependance of $J_c$ ($H//c$-axis) in Ba$_{0.6}$K$_{0.4}$Fe$_2$As$_2$ with splayed CDs introduced by (a) 2.6 GeV U ($B_\Phi$ = 3 T + 3 T, $\theta_{CD}$ = ±20°) and by (b) 320 MeV Au ($B_\Phi$ = 4 T + 4 T, $\theta_{CD}$ = ±20°) at various temperatures. Magnetic field dependence of $J_c$ at 15 K in Ba$_{0.6}$K$_{0.4}$Fe$_2$As$_2$ with splayed CDs introduced by (c) 2.6 GeV U ($B_\Phi$ = 3 T + 3 T, $\theta_{CD}$ = ±20°), by (d) 320 MeV Au ($B_\Phi$ = 4 T + 4 T, $\theta_{CD}$ = ±20°), and with tilted CDs introduced by (e) 2.6 GeV U ($B_\Phi$ = 3 T, $\theta_{CD}$ = 20°) and by (f) 320 MeV Au ($B_\Phi$ = 4 T, $\theta_{CD}$ = 20°) at various field angles.

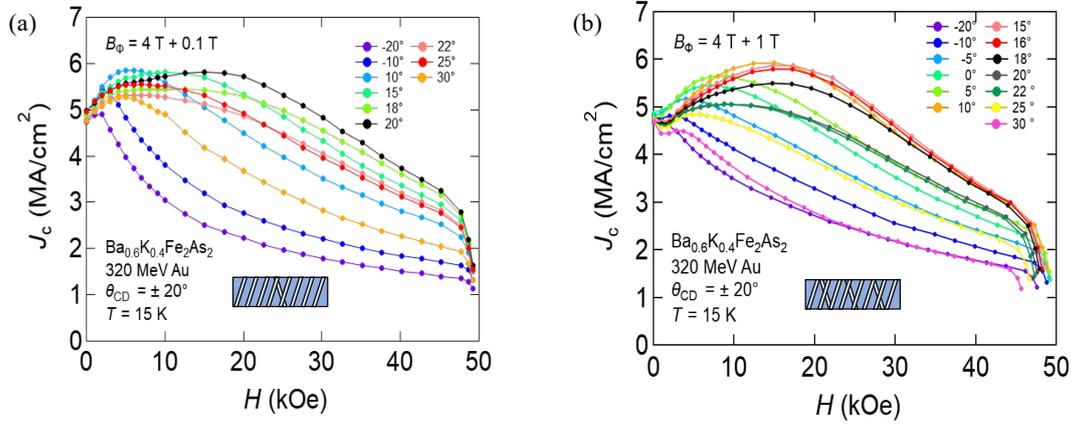

Fig. 4. (Color online) Magnetic field dependence of $J_c$ at 15 K in Ba$_{0.6}$K$_{0.4}$Fe$_2$As$_2$ with asymmetric splayed CDs introduced by 320 MeV Au (a) ($B_\Phi$ = 4 T + 0.1 T, $\theta_{CD}$ = ±20º), and (b) ($B_\Phi$ = 4 T + 1 T, $\theta_{CD}$ = ±20º).

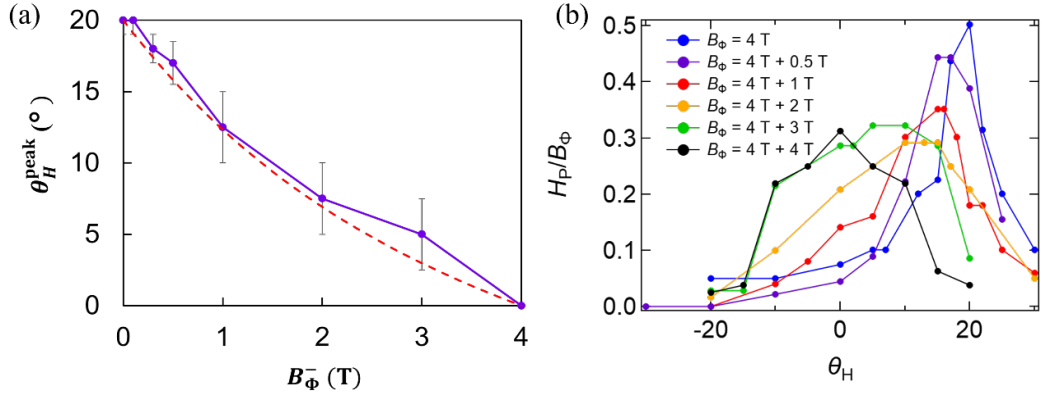

Fig. 5. (Color online) (a) The field angle, $\theta_H^{\text{peak}}$, where the peak field for APE becomes the maximum, as a function of irradiation dose along $\theta_{CD} = -20°$ ($B_\Phi^-$) in Ba$_{0.6}$K$_{0.4}$Fe$_2$As$_2$ with $B_\Phi^+ = 4$ T along $\theta_{CD} = 20°$. The red dashed line represents the plot of Eq. (6). (b) Variation of the peak field at various $\theta_H$ in crystals with different configurations of CDs. The peak field is normalized by the total matching field $B_\Phi$.

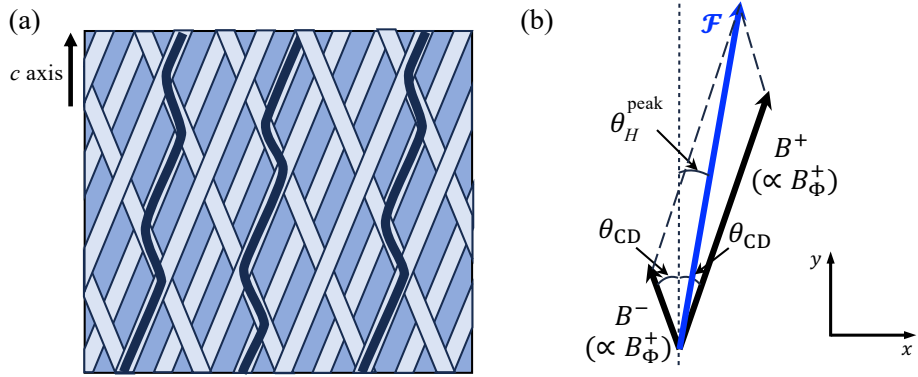

Fig. 6. (Color online) (a) Schematic image of arrangements of CDs and trapped vortices. (b) Schematic diagram showing the condition for the angle of the strongest APE, $\theta_H^{\text{peak}}$. $B^+$ and $B^-$ are internal magnetic inductions along the two directions of columnar defects.

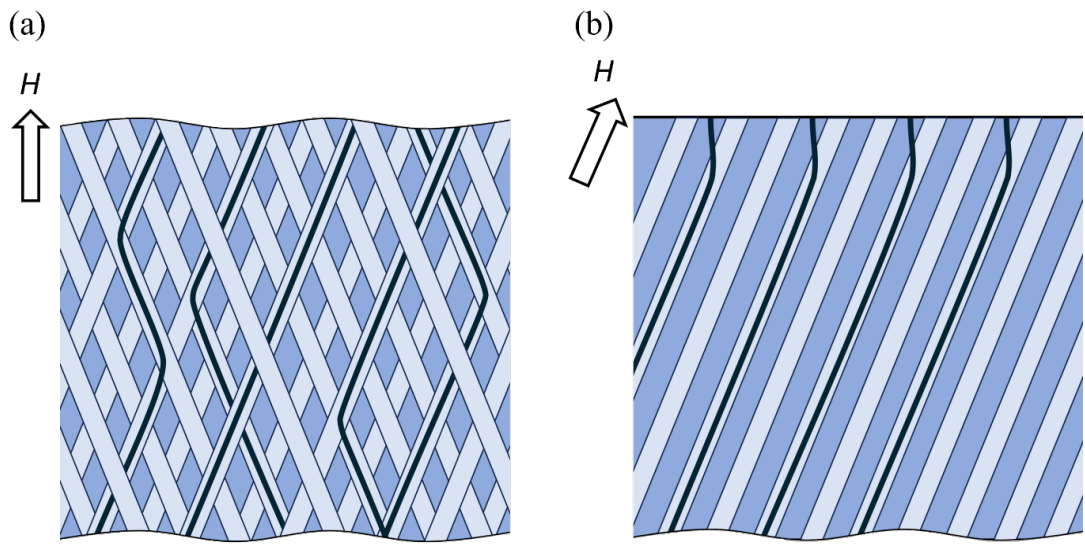

Fig. 7. (Color online) (a) Schematic image of splayed CDs and trapped vortices for $H//c$-axis. (b) Schematic image of tilted CDs and trapped vortices for $H//$CDs. Vortices form kinks near the surface of the sample.

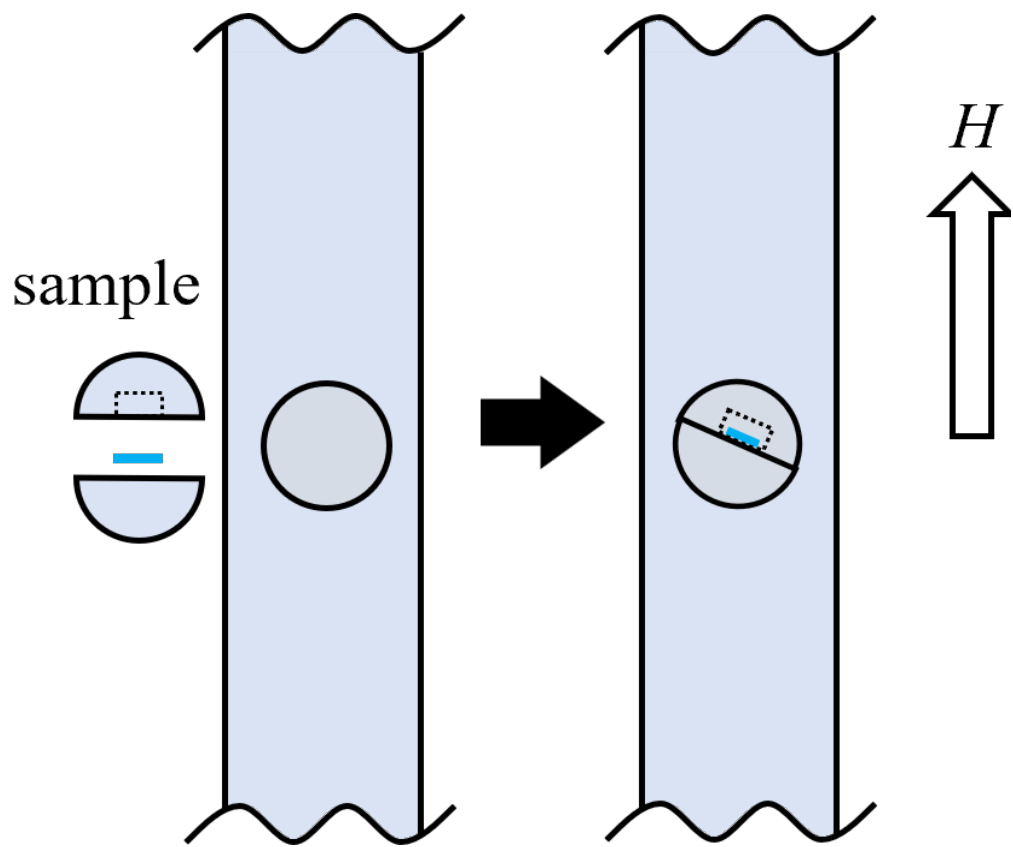

Fig. S1.   Schematic images of a rotating sample holder.

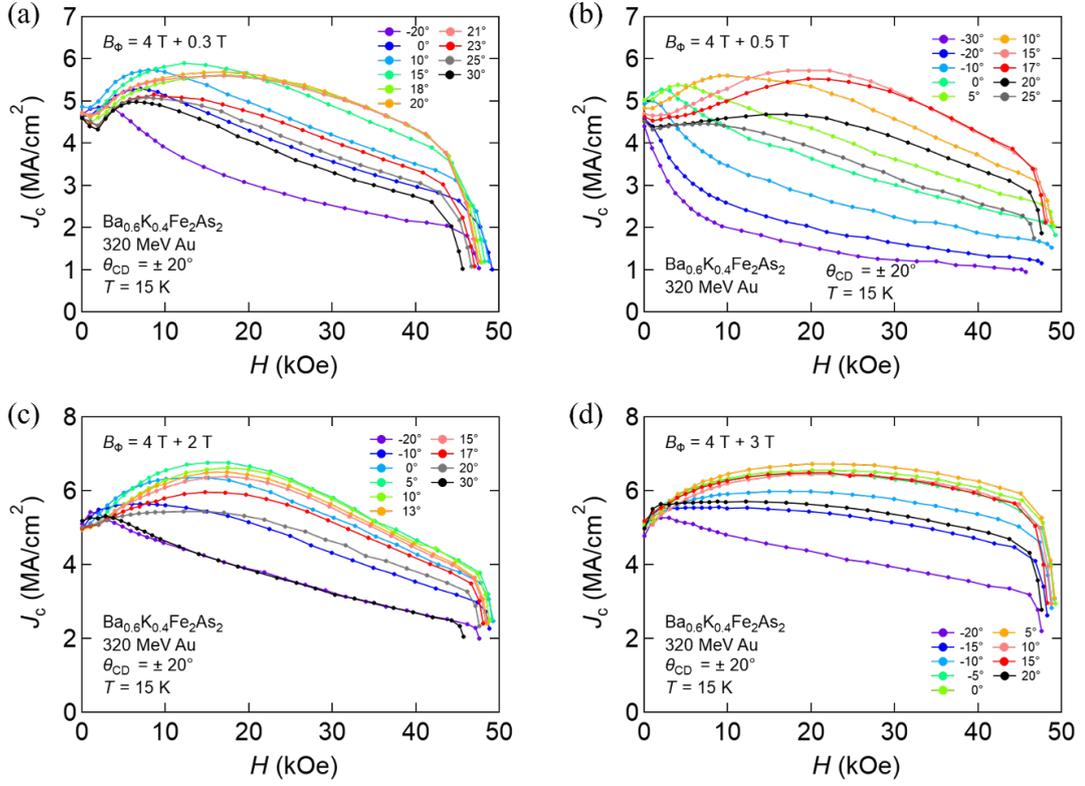

Fig. S2. Magnetic field dependence of $J_c$ at 15 K in $Ba_{0.6}K_{0.4}Fe_2As_2$ with asymmetric splayed CDs introduced by 320 MeV Au (a) ($B_\Phi = 4$ T + 0.3 T, $\theta_{CD} = \pm 20°$), (b) ($B_\Phi = 4$ T + 0.5 T, $\theta_{CD} = \pm 20°$), (c) ($B_\Phi = 4$ T + 2 T, $\theta_{CD} = \pm 20°$), and (d) ($B_\Phi = 4$ T + 3 T, $\theta_{CD} = \pm 20°$).